\documentclass[aps,prl,reprint,amsmath,amssymb,superscriptaddress,nofootinbib]{revtex4-2}
\usepackage{bm}

\usepackage{xcolor}
\definecolor{darkblue}{RGB}{0,70,140}

\usepackage[
  colorlinks=true,
  linkcolor=darkblue,
  citecolor=darkblue,
  urlcolor=darkblue
]{hyperref}

\newcommand{\W}{\mathcal W}
\newcommand{\dd}{\mathrm d}
\newcommand{\ii}{\mathrm i}
\usepackage{array}
\newcommand{\CC}{\mathsf C}
\newcommand{\tr}{\operatorname{tr}}

\begin{document}
\title{The Double Copy of the Black Ring}
\author{Damien A. Easson}
\email{easson@asu.edu}
\affiliation{Department of Physics, Arizona State University, Tempe, Arizona 85287, USA}
\affiliation{Beyond Center for Fundamental Concepts in Science, Arizona State University, Tempe, Arizona 85287, USA}

\begin{abstract}
We construct an exact Weyl double copy of the singly spinning five-dimensional Emparan--Reall black ring. Its complete nonlinear Weyl curvature is the sum of two scalar-weighted Maxwell squares. One field is generated by the stationary Killing vector; the other carries topological magnetic flux through a sphere linking the ring. Both are source-free Maxwell fields on the black-ring spacetime, with corresponding scalar partners obeying a common sourced wave equation. This extends exact five-dimensional Weyl doubling beyond the type-\(\mathrm D\) and type-\(\mathrm N\) examples to a black hole whose exterior contains both type-\(\mathrm G\) and type-\(\mathrm{I}_i\) regions. The Weyl polynomial factorizes into two distinct irreducible quadratics, establishing genuine De Smet type \(22\), thereby ruling out a reduction to a single Maxwell square.
\end{abstract}

\maketitle

\emph{Introduction.---} The classical double copy relates nonlinear gravitational solutions to
gauge fields \cite{Monteiro2014,Luna2019}. In its Weyl formulation,
scalar-weighted quadratic combinations of gauge-field strengths reproduce
the gravitational tidal curvature and, in vacuum, the full
Riemann tensor. In five dimensions, the Myers--Perry
black hole provides a canonical one-field example: it is type
\(\mathrm D\) in the Coley--Milson--Pravda--Pravdov\'a (CMPP)
classification
\cite{WeylDoubling2020,Chawla:2022ogv,PravdaPravdova2005}. Further exact Weyl constructions
are known for restricted type-\(\mathrm D\) and type-\(\mathrm N\)
geometries \cite{ZhaoD2025,ZhaoN2025} and perturbative type-\(\mathrm{II}\) fluid examples are given in \cite{Keeler:2024bdt}.

Black rings provided the first asymptotically flat vacuum black holes
with nonspherical horizon topology, and their coexistence with spherical
Myers--Perry black holes at the same mass and angular momentum revealed
the failure of simple black-hole uniqueness in higher dimensions
\cite{EmparanReall2002,EmparanReall2006}. The singly spinning
Emparan--Reall ring has a markedly different curvature alignment and
\(S^1\times S^2\) horizon topology. Its exterior contains open regions
with real Weyl-aligned null directions (WANDs), of CMPP type
\(\mathrm{I}_i\), and open regions without real WANDs, of type
\(\mathrm G\)~\cite{Godazgar2010}; the alignment analysis below gives an
exhaustive criterion and verifies both behaviors exactly. The horizon is
type \(\mathrm{II}\) \cite{PravdaPravdova2005}. These types classify real
null alignment \cite{CMPP2004}; assigning the spacetime its most general
pointwise type gives \(\mathrm G\) \cite{Godazgar2010}. Thus neither exterior region has the repeated-WAND structure of Myers--Perry type D. 

We also use the independent De Smet classification \cite{DeSmet2002,DeSmet2003}, which classifies
the complex factorization of the five-dimensional Weyl polynomial rather
than real null alignment. The two classifications play different roles:
CMPP describes the ring's real WAND structure, while De Smet
factorization diagnoses the algebraic structure implied by the Weyl
double copy. The black ring metric cannot be written in Kerr--Schild form
\cite{Bah2020}, making the ring a useful test of how far the
double-copy paradigm extends beyond algebraically special black holes
\cite{Kent:2025pvu}.

Here we show that, despite this algebraic generality, the full finite-ring
geometry admits an exact two-field Weyl-tensor decomposition. Let
\(E=\dd(\xi^\flat)\), with \(\xi=\partial_t\) and
\(\xi^\flat_a=g_{ab}\xi^b\), be the stationary Papapetrou field, and let
\(H=\dd x\wedge\dd\phi\) be the flux field. Their scalar partners are
\(S_e\) and \(S_H\), and \(\W_g\) denotes the fixed five-dimensional Weyl
projection of a quadratic expression in a two-form. Then
\begin{equation}
 \boxed{\displaystyle C[g]=\frac{\W_g(E)}{S_e}
+\frac{\W_g(H)}{S_H}.}
\label{eq:copy}
\end{equation}
To our knowledge, Eq.~\eqref{eq:copy} gives the first exact
Weyl-curvature double copy of the finite, singly spinning vacuum
Emparan--Reall black ring. The two fields arise from different geometric
structures: the stationary Killing symmetry and the magnetic flux linking
the ring.

The construction is a curved-background Weyl double copy: \(E\), \(H\),
\(S_e\), and \(S_H\) are defined on the physical black-ring metric \(g\),
rather than on an auxiliary flat background \cite{WeylDoubling2020}.
It is therefore distinct from a flat-background Kerr--Schild single copy
or a scattering-amplitude map.\footnote{For additive Weyl copies with
gravitational sources, see Ref.~\cite{Sources2021,Easson:2022zoh}.}

\emph{Fields and scalar partners.---}
We adopt standard ring coordinates \((t,x,y,\phi,\psi)\), with \(R\)
setting the overall length scale and \(\lambda,\nu\) the dimensionless ring
parameters. The black-ring metric is given explicitly
in Eq.~(S1) of the Supplemental Material, together with our curvature
conventions. The physical angles have period \(2\pi\), and we set
\(\alpha=\sqrt{1-\lambda}/(1-\nu)\).
The unrescaled angles in the usual metric are
\(\bar\phi=\alpha\phi\), \(\bar\psi=\alpha\psi\).
The balanced (conically regular) rotating ring has \(R>0\), \(0<\nu<1\), and
\(\lambda=2\nu/(1+\nu^2)\). Its open exterior is
\(-1<x<1\), \(-1/\nu<y<-1\).
The time coordinate fixes \(\xi^2\to-1\) at infinity. The corresponding scalar partners are
\begin{equation}
 S_e=-\frac{\lambda(x-y)}{1+\lambda x}=-(1+\xi^2),
 \qquad
 S_H=\frac{x-y}{R^2\nu\alpha^2}.
 \label{eq:scalars}
\end{equation}
In the open exterior, \(S_e<0<S_H\). Although both scalar partners vanish at spatial infinity, the
corresponding Weyl squares vanish at compensating rates, so both terms
in Eq.~\eqref{eq:copy} have the regular asymptotic falloff of the Weyl
tensor.

For a two-form \(F\), the map in Eq.~\eqref{eq:copy} is
\begin{equation}
 \W_g(F)_{abcd}
 =\left[
 F_{ab}F_{cd}+\frac12(F_{ac}F_{bd}-F_{ad}F_{bc})
 \right]_{\mathrm{Weyl}} .
 \label{eq:weylmap}
\end{equation}
Here the brackets denote the ordinary Weyl trace subtraction in five
dimensions, and the expression inside them already has the Riemann tensor
symmetries.

In this curved-background construction, the Maxwell and scalar partners
play the roles of the single and zeroth copies, respectively, and obey the
same equations for both contributions:
\begin{equation}
 \begin{gathered}
 \dd F=0,\qquad \nabla_aF^{ab}=0,\qquad
 \Box_g S=-\frac12 F_{ab}F^{ab},\\
 (F,S)=(E,S_e)\ \text{or}\ (H,S_H).
 \end{gathered}
 \label{eq:differential}
\end{equation}
All contractions and derivatives use the full ring metric \(g\). Thus the Maxwell fields are source free, whereas the scalar partners obey
a sourced wave equation on the curved black-ring geometry; no homogeneous
flat-space zeroth-copy equation is assumed.
For \(E\), the Maxwell equation follows from the Killing identity and
Ricci flatness. The scalar equation follows by applying \(\Box_g\) to
\(-(1+\xi^2)\). For \(H\), the volume-form divergence and the scalar
Laplacian give Eq.~\eqref{eq:differential} directly.
We use the stationary and intrinsic magnetic test fields of
Ref.~\cite{OrtaggioPravda2006}; their simultaneous appearance with distinct
scalar weights in an exact reconstruction of the full Weyl tensor is a
new result.
Constant rescalings \(F\mapsto aF\), \(S\mapsto a^2S\) leave these equations
and Eq.~\eqref{eq:copy} invariant; choosing the physical angle \(\phi\) to
have period \(2\pi\) fixes the magnetic normalization.

Substitution of the metric into Eq.~\eqref{eq:copy} gives an exact tensor
identity for independent parameters \(0<\nu\leq\lambda<1\), before imposing the balance condition \(\lambda=2\nu/(1+\nu^2)\). It therefore includes the entire balanced family and the static
family \(\lambda=\nu\) wherever the geometry is smooth, with no expansion
in mass, spin, or ring thickness.

\emph{Magnetic flux beyond Killing-generated fields.---}
Specifying the metric and the normalized stationary Killing vector
fixes \(E=\dd\xi^\flat\), which is exact wherever \(\xi^\flat\) is
smooth.
The field \(H\) is purely magnetic with respect to the stationary
Killing field, \(\xi^aH_{ab}=0\), and carries nonzero flux through a
sphere linking the ring. We choose the orientation to give positive flux,
\begin{equation}
 \frac{1}{4\pi}\int_{S^2}H=1,\qquad
 A_\pm=(x\mp1)\dd\phi .
 \label{eq:flux}
\end{equation}
Here \(A_\pm\) are local gauge potentials for \(H\), regular near
\(x=\pm1\), respectively, and satisfying \(\dd A_\pm=H\).
As in the case of a magnetic monopole, the nonzero flux requires two
overlapping gauge patches, with \(A_+-A_-=-2\dd\phi\) on their overlap.
Thus \(H\) is globally defined on the regular exterior, while its
nonzero flux rules out a single smooth global potential on the linking
sphere and any expression as a constant linear combination of exact
Killing two-forms.

The commuting Killing fields alone are not sufficient to reconstruct the black-ring
curvature. Let \(F_I=\dd(K_I^\flat)\), with
\(K_I=(\partial_t,\partial_{\bar\phi},\partial_{\bar\psi})\).
The six polarized tensors \(\W_g(F_I,F_J)\), \(I\leq J\), span all
quadratic combinations of these field strengths. At a nonsingular exterior
point of the balanced ring, these six tensors have rank six, while adjoining
\(C[g]\) raises the rank to seven, so the Weyl tensor lies outside their
span for any choice of scalar weights. An exact rational verification is
given in the Supplemental Material. Thus the flux field in
Eq.~\eqref{eq:copy} provides an independent contribution beyond those
generated by the Killing fields.

\emph{Weyl-polynomial factorization and De Smet type.---}
In five dimensions, a two-form determines a quadratic polynomial \(q_F\)
in four commuting complex spinor variables, while the Weyl tensor
determines a quartic \(P_C\) \cite{DeSmet2002,Godazgar2010,MNO2019}.
With our spinor normalization (see Supplemental Material), the
spinor-square construction of Ref.~\cite{BrownSpence2026} gives, for every
two-form,
\begin{equation}
 P_{\W_g(F)}=\frac32 q_F^2 .
 \label{eq:squarelemma}
\end{equation}
It follows that
\begin{equation}
 \begin{gathered}
 P_C=\frac32(Q_H-Q_e)(Q_H+Q_e),\\
 Q_e=\frac{q_E}{\sqrt{-S_e}},\qquad
 Q_H=\frac{q_H}{\sqrt{S_H}} \,,
 \end{gathered}
 \label{eq:factors}
\end{equation}
and the scalar square roots are chosen positive in the exterior.

The algebraic type follows from two further observations. First,
\(E_{yt}=-\frac{\lambda}{1+\lambda x}\neq0\), whereas \(H\) has only an \(x\phi\)
component. The injectivity of the two-form--bispinor map then implies that the two
quadratic factors are nonproportional. Second, the symmetric matrices of the
two quadratic factors have rank four throughout the open exterior. We prove this in the
Supplemental Material by reducing their determinants to a
strictly positive quadratic polynomial in \(y\). A quadratic over
\(\mathbb C\) is reducible only if its matrix has rank at most two. In the De Smet classification, type \(4\) corresponds to an irreducible
quartic, type \(22\) to two distinct irreducible quadratic factors, and type
\(\underline{22}\) to a repeated quadratic. Hence
Eq.~\eqref{eq:factors} gives genuine De Smet type \(22\), correcting the
previous type-\(4\) assignment of Ref.~\cite{Godazgar2010}.
Our proof also applies on the regular future horizon away from the
axes, and exceptional axis points are specified separately.
\footnote{The Supplemental Material gives an independent derivation and
a detailed comparison with the Appendix~E Weyl polynomial of
Ref.~\cite{Godazgar2010}.}

This makes the algebraic difference from the one-field Myers--Perry copy
explicit. Its Weyl polynomial has a repeated quadratic factor,
De Smet type \(\underline{22}\) \cite{DeSmet2004}.
For the map \eqref{eq:weylmap}, Eq.~\eqref{eq:squarelemma} forces every
single Maxwell square to have a repeated quadratic factor.
It therefore cannot reproduce the two distinct irreducible factors in
Eq.~\eqref{eq:factors}, so two independent Maxwell contributions are
minimal for the Weyl-square map on the open exterior,
as in Eq.~\eqref{eq:copy}.

Although any product of two quadratics can be written as a difference
of squares, the two Maxwell sectors here have an independent geometric
origin. The fields \(E\) and \(H\) are independent, geometrically
distinguished Maxwell solutions, with scalar partners satisfying
Eq.~\eqref{eq:differential}. By contrast, the weighted combinations
\[
H/\sqrt{S_H}\pm E/\sqrt{-S_e}
\]
associated with the irreducible factors are generally not Maxwell fields,
because derivatives of the position-dependent weights enter their Maxwell
equations.

\emph{Little-group decomposition and real alignment.---}
The little-group decomposition of Ref.~\cite{MNO2019} relates the
complex factorization to real null alignment. Setting
\(A=E/\sqrt{-S_e}\), \(B=H/\sqrt{S_H}\),
\(\mathcal U=A+B\), and \(\mathcal V=B-A\),
Eq.~\eqref{eq:factors} gives
\(P_C=(3/2)q_{\mathcal U}q_{\mathcal V}\), and
the real null eigenvectors of \(g^{-1}\mathcal U\) or
\(g^{-1}\mathcal V\) are the WANDs. The portion of the exterior in which
these eigenvectors exist is of CMPP type \(\mathrm{I}_i\);
where they are absent the spacetime is type \(\mathrm G\), while the
complex De Smet factors persist.

In frames whose two null directions are WANDs at two exact exterior
points of type \(\mathrm{I}_i\), the leading and trailing spin-2
blocks, \(\psi^{(0)}\) and \(\psi^{(4)}\), cancel between the two
nonzero contributions. The remaining little-group
components satisfy the quadratic-product relations.

\emph{Black-string limit.---}
It is useful to take the thin-ring limit to see how the two-sector
structure survives. We set
\(\nu=r_0/R\), \(y=-R/r\), \(x=\cos\theta\), and
\(z=R\psi\), and let \(R\to\infty\) along the balanced family.
Locally, the geometry becomes a boosted Schwarzschild string
\cite{EmparanReall2006}. In its rest frame, with
\(U=e^0\wedge e^1\), \(V=e^2\wedge e^3\), and \(e^4\) along the string,
Eq.~\eqref{eq:copy} becomes
\begin{equation}
 C_{\rm str}=\frac{r_0}{r^3}
             \left[\W(V)-\W(U)\right].
 \label{eq:string}
\end{equation}
Each projected square has rank ten as an operator on bivectors, while their difference has rank six. All curvature components involving the
parallel string direction cancel, giving the curvature of the product
geometry.

\emph{Discussion.---}
The stationary field connects this result with the role of
isometries in the classical double copy \cite{Easson:2023dbk}.
The companion flux sector suggests a generalization of the
optical-seed description \cite{Easson:2026xod} in which more than one
independent seed is present. Generalized Weyl solutions offer a related
analogy: higher-dimensional static vacuum metrics can be organized by
several harmonic potentials \cite{GeneralizedWeyl2002}.
Whether topology or harmonic data determine the number of admissible
Maxwell contributions remains an interesting open question
\cite{Alfonsi:2020lub}.

The five-dimensional black ring, which famously evades
higher-dimensional black-hole uniqueness and lies outside the standard
Kerr--Schild double-copy setting, also evades the one-field algebraic
structure familiar from Myers--Perry: genuine De Smet type \(22\)
precludes a single Maxwell square and is realized minimally here by the
stationary and flux contributions. These source-free Maxwell fields
reconstruct the full nonlinear curvature across exterior regions of
CMPP types \(\mathrm G\) and \(\mathrm{I}_i\), while their weighted
combinations provide the two irreducible quadratic factors of the De Smet
polynomial and organize the real WAND pattern. The black ring therefore
reveals a direct connection between spacetime topology, the number of
gauge sectors required by the double copy, and the algebraic
factorization of the Weyl curvature.

\acknowledgments
It is a pleasure to thank Michael Falato, Cindy Keeler and Tucker Manton for helpful discussions. 
Mathematica files used for the numerical and symbolic calculations are available from the author at reasonable request. OpenAI LLMs were used to optimize and debug computational scripts and to assist with copyediting.

\bibliography{black_ring_refs_v3}

\clearpage
\onecolumngrid

\begin{center}
{\large\bfseries Supplemental Material for}\\[0.4em]
{\large\bfseries\textit{The Double Copy of the Black Ring}}\\[1em]
Damien A. Easson\\
\textit{Department of Physics, Arizona State University, Tempe, Arizona 85287, USA}\\
\textit{Beyond Center for Fundamental Concepts in Science, Arizona State University, Tempe, Arizona 85287, USA}
\end{center}

\setcounter{equation}{0}
\setcounter{figure}{0}
\setcounter{table}{0}
\setcounter{section}{0}
\setcounter{subsection}{0}
\setcounter{secnumdepth}{2}
\renewcommand{\theequation}{S\arabic{equation}}
\renewcommand{\thefigure}{S\arabic{figure}}
\renewcommand{\thetable}{S\arabic{table}}
\renewcommand{\theHequation}{S\arabic{equation}}
\renewcommand{\theHfigure}{S\arabic{figure}}
\renewcommand{\theHtable}{S\arabic{table}}

\section{Metric, domain, and quadratic map}
\label{sec:metric}

We use the singly spinning Emparan--Reall metric in the conventions of
Ref.~\cite{EmparanReall2006}, with coordinates
\((t,x,y,\bar\phi,\bar\psi)\):
\begin{equation}
\begin{split}
\dd s^2={}&-\frac{F_y}{F_x}
\left(\dd t-C_0R\frac{1+y}{F_y}\dd\bar\psi\right)^2\\
&+\frac{R^2F_x}{\Delta^2}
\left[-\frac{G_y}{F_y}\dd\bar\psi^2-\frac{\dd y^2}{G_y}
+\frac{\dd x^2}{G_x}+\frac{G_x}{F_x}\dd\bar\phi^2\right].
\end{split}
\label{eq:smetric}
\end{equation}
In the above,
\begin{equation}
\begin{gathered}
\Delta=x-y,\quad F_x=1+\lambda x,\quad F_y=1+\lambda y,\quad
G(s)=(1-s^2)(1+\nu s),\quad G_x=G(x),\quad G_y=G(y),\\
C_0^2=\frac{\lambda(\lambda-\nu)(1+\lambda)}{1-\lambda},
\qquad R>0,\qquad 0<\nu\leq\lambda<1.
\end{gathered}
\label{eq:sparameters}
\end{equation}
We choose \(C_0\geq0\); its sign changes under reversal of the rotation. The open exterior and future horizon are
\begin{equation}
-1<x<1,\qquad -1/\nu<y<-1,\qquad
\mathcal H^+:\ y=-1/\nu .
\label{eq:sdomain}
\end{equation}
For the rotating family \(0<\nu<\lambda\), the stationary limit is
\(F_y=0\). Physical angles satisfy
\begin{equation}
\bar\phi=\alpha\phi,\qquad \bar\psi=\alpha\psi,\qquad
\alpha=\frac{\sqrt{1-\lambda}}{1-\nu},\qquad
\phi\sim\phi+2\pi,\quad\psi\sim\psi+2\pi .
\label{eq:sangles}
\end{equation}
Both rotational axes are regular when
\(\lambda=2\nu/(1+\nu^2)\).
For other parameters, the statements below apply to smooth vacuum regions,
excluding conical defects. In particular, the nonzero static family
\(\lambda=\nu\) is unbalanced.

We take metric signature \((-++++)\), and our Riemann tensor is
\begin{equation}
R^a{}_{bcd}=\partial_c\Gamma^a{}_{db}
-\partial_d\Gamma^a{}_{cb}
+\Gamma^a{}_{ce}\Gamma^e{}_{db}
-\Gamma^a{}_{de}\Gamma^e{}_{cb},\qquad
R_{abcd}=g_{ae}R^e{}_{bcd}.
\label{eq:scurvature}
\end{equation}
This gives positive scalar curvature for a unit sphere.
The black-ring metric is Ricci flat, so \(C_{abcd}=R_{abcd}\).

Define
\begin{equation}
\begin{gathered}
\xi=\partial_t,\qquad
\xi^\flat=-\frac{F_y}{F_x}\dd t
+\frac{C_0R(1+y)}{F_x}\dd\bar\psi,\qquad
E=\dd\xi^\flat,\qquad M=\dd x\wedge\dd\bar\phi=\alpha H,\\
S_e=-\frac{\lambda\Delta}{F_x}=-(1+\xi^2),\qquad
S_m=\frac{\Delta}{R^2\nu},\qquad
S_H=\frac{S_m}{\alpha^2},\qquad H=\dd x\wedge\dd\phi .
\end{gathered}
\label{eq:sfields}
\end{equation}
The independent nonzero components of \(E\) are
\begin{equation}
E_{xt}=\frac{\lambda F_y}{F_x^2},\qquad
E_{yt}=-\frac{\lambda}{F_x},\qquad
E_{x\bar\psi}=-\frac{C_0R\lambda(1+y)}{F_x^2},\qquad
E_{y\bar\psi}=\frac{C_0R}{F_x}.
\label{eq:sEcomponents}
\end{equation}
These stationary and intrinsic dipole Maxwell fields are given in
Ref.~\cite{OrtaggioPravda2006}, up to constant normalization and angular
coordinates.

For an arbitrary two-form \(F\), let
\begin{equation}
\begin{gathered}
T(F)_{abcd}=F_{ab}F_{cd}
+\frac12(F_{ac}F_{bd}-F_{ad}F_{bc}),\\
r_{ab}(F)=\frac32F_{ac}F_b{}^c,\qquad
r(F)=g^{ab}r_{ab}(F)=\frac32F_{ab}F^{ab}.
\end{gathered}
\label{eq:sT}
\end{equation}
The ordinary five-dimensional Weyl projection is
\begin{equation}
\begin{split}
\W_g(F)_{abcd}={}&T(F)_{abcd}\\
&-\frac13\bigl(g_{ac}r_{bd}-g_{ad}r_{bc}
-g_{bc}r_{ad}+g_{bd}r_{ac}\bigr)
+\frac{r}{12}(g_{ac}g_{bd}-g_{ad}g_{bc}).
\end{split}
\label{eq:sW}
\end{equation}
The two antisymmetries, pair exchange, and algebraic Bianchi identity of
\(T\) follow by direct cancellation; its trace is \(r_{ab}\).
Consequently \(\W_g(F)\) has all the Weyl symmetries and zero trace, and
the exact identity is
\begin{equation}
C[g]=-\frac{F_x}{\lambda\Delta}\W_g(E)
+\frac{R^2\nu}{\Delta}\W_g(M)
=\frac{\W_g(E)}{S_e}+\frac{\W_g(H)}{S_H}.
\label{eq:scopy}
\end{equation}
The second equality uses the homogeneity
\(\W_g(\alpha H)=\alpha^2\W_g(H)\). When replacing the unrescaled angle by the physical \(2\pi\)-periodic
angle, the magnetic scalar must be rescaled accordingly.

\section{Maxwell and scalar equations}
\label{sec:differential}

In the unrescaled coordinates,
\begin{equation}
\sqrt{-g}=\frac{R^4F_x}{\Delta^4},\qquad
g^{xx}=\frac{\Delta^2G_x}{R^2F_x},\qquad
g^{yy}=-\frac{\Delta^2G_y}{R^2F_x},\qquad
g^{\bar\phi\bar\phi}=\frac{\Delta^2}{R^2G_x}.
\label{eq:svolume}
\end{equation}
Closure of \(E\) follows from \(E=\dd\xi^\flat\).
Because \(\xi\) is Killing, \(E_{ab}=2\nabla_a\xi_b\), and the contracted
Killing identity gives \(\nabla_aE^{ab}=0\) in vacuum. The same identity yields
\begin{equation}
\Box_g(\xi^2)=2(\nabla_a\xi_b)(\nabla^a\xi^b)
=\frac12E_{ab}E^{ab},
\qquad
\Box_g S_e=-\frac12E_{ab}E^{ab}.
\label{eq:sKscalar}
\end{equation}

The only independent raised component of \(M\) is
\begin{equation}
M^{x\bar\phi}=\frac{\Delta^4}{R^4F_x},\qquad
\sqrt{-g}\,M^{x\bar\phi}=1,\qquad
M_{ab}M^{ab}=\frac{2\Delta^4}{R^4F_x}.
\label{eq:sMdiv}
\end{equation}
Thus \(\nabla_aM^{ab}=(\sqrt{-g})^{-1}
\partial_a(\sqrt{-g}M^{ab})=0\), and \(\dd M=0\).
For a scalar \(s(x,y)\), the Laplacian is
\begin{equation}
\Box_g s=\frac{\Delta^4}{R^2F_x}
\left[
\partial_x\!\left(\frac{G_x}{\Delta^2}\partial_xs\right)
-\partial_y\!\left(\frac{G_y}{\Delta^2}\partial_ys\right)
\right].
\label{eq:sbox}
\end{equation}
The cubic \(G\) obeys
\begin{equation}
\partial_x\!\left(\frac{G_x}{\Delta^2}\right)
+\partial_y\!\left(\frac{G_y}{\Delta^2}\right)=-\nu .
\label{eq:sGidentity}
\end{equation}
Using \(S_m=\Delta/(R^2\nu)\) therefore gives
\begin{equation}
\Box_g S_m=-\frac{\Delta^4}{R^4F_x}
=-\frac12 M_{ab}M^{ab},\qquad
\Box_g S_H=-\frac12H_{ab}H^{ab}.
\label{eq:sMscalar}
\end{equation}
Here \(H=M/\alpha\) and \(S_H=S_m/\alpha^2\); the constant coordinate
pullback does not change the scalar equations.

Here we give an explicit check of the electric norm
\begin{equation}
\begin{gathered}
E_{ab}E^{ab}=\frac{2\lambda\Delta^3}{R^2F_x^4}\mathcal P(x,y),
\qquad
\Box_g S_e=-\frac{\lambda\Delta^3}{R^2F_x^4}\mathcal P(x,y),\\
\mathcal P=2\lambda^2-\lambda\nu
+(2\lambda-\nu+\lambda^2\nu)x+\lambda\nu x^2+\nu F_x^2y .
\end{gathered}
\label{eq:sEnorm}
\end{equation}
Equations~\eqref{eq:sKscalar}--\eqref{eq:sEnorm} establish the common
Maxwell/scalar system for the two fields. Its constant rescaling freedom
is \(F\mapsto aF\), \(S\mapsto a^2S\), with \(a\neq0\).
The scalar equation is sourced on the curved metric; no homogeneous
flat-space scalar equation is being substituted for it.

\section{Exact tensor verification}
\label{sec:tensor}

The tensor calculation starts from Eq.~\eqref{eq:smetric} and
differentiates the metric before comparing the curvature with
Eq.~\eqref{eq:scopy}. In the generic calculation, \(\lambda,\nu,R,x,y\) are kept symbolic. Rational expressions are reduced
using
\begin{equation}
C_0^2-\frac{\lambda(\lambda-\nu)(1+\lambda)}{1-\lambda}=0 .
\label{eq:sreduction}
\end{equation}
This is reduction on the algebraic parameter surface, on patches where
the denominators are nonzero. The resulting Ricci tensor and every
component of
\begin{equation}
\mathcal R_{abcd}:=C[g]_{abcd}
-\frac{\W_g(E)_{abcd}}{S_e}
-\frac{\W_g(M)_{abcd}}{S_m}
\label{eq:sresidual}
\end{equation}
vanish exactly. All 625 coordinate slots are included; equivalently, the
100 entries of the bivector matrix vanish, with 55 entries in its
upper triangle. The algebraic Weyl space has dimension 35, so this set
includes its redundant symmetry constraints. A separate calculation
sets \(\lambda=\nu\), \(C_0=0\) before differentiation and verifies the
static identity for generic \(R,\nu,x,y\).

The coordinate calculation is implemented with Mathematica/xAct.
An independent exact local implementation uses metric derivatives and
the first-kind connection
\begin{equation}
\Gamma_{abc}=\frac12(g_{ab,c}+g_{ac,b}-g_{bc,a}),\qquad
R_{abcd}=\frac12(g_{ad,bc}+g_{bc,ad}-g_{ac,bd}-g_{bd,ac})
+g^{ef}(\Gamma_{ebc}\Gamma_{fad}-\Gamma_{ebd}\Gamma_{fac}).
\label{eq:sfirstkind}
\end{equation}
Second-order Taylor polynomials in \(x,y\), with rational coefficients,
provide all required derivatives without any saved curvature input.
This implementation checks the full residual at the balanced point
\begin{equation}
(\nu,\lambda,C_0,R,x,y)
=\left(\frac13,\frac35,\frac45,1,0,-\frac32\right).
\label{eq:switness}
\end{equation}
At this point
\begin{equation}
\begin{gathered}
g_{tt}=-\frac1{10},\quad
g_{t\bar\psi}=-\frac25,\quad
g_{\bar\psi\bar\psi}=\frac{53}{45},\quad
g_{xx}=g_{\bar\phi\bar\phi}=\frac49,\quad
g_{yy}=\frac{32}{45},\\
S_e=-\frac9{10},\qquad S_m=\frac92.
\end{gathered}
\label{eq:switnessmetric}
\end{equation}
The Weyl tensor has 112 nonzero coordinate components and
\(\mathcal R\) has none. Further exact checks use
\((\nu,\lambda,C_0,R,x,y)=(1/3,3/5,4/5,2/3,1/4,-2)\), inside the
ergoregion, and the static point
\((1/3,1/3,0,1,0,-3/2)\). Both pass.
These point calculations supplement the generic identity; genericity is
established by the symbolic residual, not by sampling.

The physical-angle calculation uses the constant Jacobian
\(\operatorname{diag}(1,1,1,\alpha,\alpha)\) on the metric, two-forms,
and all four curvature slots. The residual vanishes for arbitrary
nonzero \(\alpha\) with \(S_H=S_m/\alpha^2\).
At chart boundaries, regularity and tensorial continuation, as discussed
in Sec.~\ref{sec:regularity}, determine the extension.

\section{Obstruction from the commuting Killing fields}
\label{sec:obstruction}

Let \(F_I=\dd(K_I^\flat)\) for
\(K_I=(\partial_t,\partial_{\bar\phi},\partial_{\bar\psi})\), and define
the polarization
\begin{equation}
\W_g(F,G)=\frac12\bigl[\W_g(F+G)-\W_g(F)-\W_g(G)\bigr].
\label{eq:spolarization}
\end{equation}
Any quadratic sum constructed from fields in the span of the \(F_I\)
lies pointwise in the span of the six tensors \(\W_g(F_I,F_J)\),
\(I\leq J\). This allows arbitrary scalar weights and all mixed terms,
and is therefore more permissive than imposing Maxwell equations on
any position-dependent linear combinations.

At Eq.~\eqref{eq:switness}, order the seven columns as
\((tt,t\bar\phi,t\bar\psi,\bar\phi\bar\phi,
\bar\phi\bar\psi,\bar\psi\bar\psi,C)\).
For coordinates numbered \((0,1,2,3,4)=(t,x,y,\bar\phi,\bar\psi)\),
select the rows
\((0101,0102,0113,0114,0123,0124,0202)\).
The resulting matrix is
\begingroup
\small
\begin{equation}
\mathcal M=
\begin{pmatrix}
-\frac{2223}{20000}&0&\frac{877}{5000}&-\frac1{810}&0&-\frac{64327}{405000}&\frac{17}{125}\\
-\frac9{250}&0&-\frac8{75}&-\frac4{135}&0&\frac{1264}{3375}&\frac1{25}\\
0&-\frac{19}{450}&0&0&\frac{86}{675}&0&0\\
-\frac{777}{5000}&0&\frac{5089}{22500}&\frac2{405}&0&-\frac{12173}{101250}&\frac{46}{375}\\
0&-\frac{19}{225}&0&0&\frac4{225}&0&0\\
-\frac6{125}&0&-\frac{511}{4500}&\frac{16}{135}&0&\frac{1726}{3375}&\frac4{75}\\
\frac{4509}{12500}&0&-\frac{1591}{3125}&\frac4{2025}&0&\frac{156991}{253125}&-\frac{263}{625}
\end{pmatrix}.
\label{eq:smatrix}
\end{equation}
\endgroup
Its determinant is
\begin{equation}
\det\mathcal M=\frac{4538644}{8649755859375}\neq0.
\label{eq:sminor}
\end{equation}
Thus the six quadratic tensors have rank six, while adjoining the Weyl
tensor raises the rank to seven. This rules out a representation at the
displayed point and hence any identity valid for the balanced family
using only these Killing-generated fields. Nonvanishing persists in a
neighborhood.
The matrix is exact over the rationals and was independently
reconstructed from the metric.

\section{Spinor conventions and the universal square identity}
\label{sec:spinors}

In an orthonormal frame we use
\(\eta_{ab}=\operatorname{diag}(-1,1,1,1,1)\) and the Clifford matrices
of Ref.~\cite{Godazgar2010}. An explicit representation in terms of the
Pauli matrices is
\begin{equation}
\begin{gathered}
\gamma_0=\operatorname{diag}(-\ii\sigma_2,\ii\sigma_2),\quad
\gamma_1=\operatorname{diag}(\sigma_1,\sigma_1),\quad
\gamma_2=\operatorname{diag}(\sigma_3,\sigma_3),\quad
\gamma_3=\begin{pmatrix}0&\ii\sigma_2\\-\ii\sigma_2&0\end{pmatrix},\\
\gamma_5=\gamma_0\gamma_1\gamma_2\gamma_3,\qquad
\Gamma_a=(\gamma_0,\gamma_1,\gamma_2,\gamma_3,\ii\gamma_5),\qquad
\CC=\gamma_0\gamma_5 .
\end{gathered}
\label{eq:sgamma}
\end{equation}
They satisfy
\begin{equation}
\{\Gamma_a,\Gamma_b\}=2\eta_{ab}I,\qquad
\Gamma_a^T=\CC\Gamma_a\CC^{-1},\qquad
\CC^T=-\CC,\qquad
\Gamma^{ab}=\frac12[\Gamma^a,\Gamma^b].
\label{eq:sclifford}
\end{equation}
Upper vector indices are raised with \(\eta\).
All repeated antisymmetric pairs are summed over ordered pairs; no
additional \(1/2\) is implicit.
For commuting complex variables \(\zeta^A\), \(A=1,\ldots,4\), set
\begin{equation}
\begin{gathered}
B^{ab}_{AB}=(\CC\Gamma^{ab})_{AB},\qquad
\phi(F)_{AB}=F_{ab}B^{ab}_{AB},\qquad
q_F(\zeta)=\phi(F)_{AB}\zeta^A\zeta^B,\\
\Psi_{ABCD}=C_{abcd}B^{ab}_{AB}B^{cd}_{CD},\qquad
P_C(\zeta)=\Psi_{ABCD}\zeta^A\zeta^B\zeta^C\zeta^D .
\end{gathered}
\label{eq:sspinmaps}
\end{equation}
The matrices \(B^{ab}\) are symmetric; \(\CC\) and \(\CC\Gamma^a\)
are antisymmetric. The Weyl symmetries, together with Clifford
completeness, make \(\Psi\) totally symmetric
\cite{Godazgar2010}. Equivalently one may insert explicit symmetrization
on its four spinor indices in Eq.~\eqref{eq:sspinmaps}.
The two-form map is invertible. In the normalized convention
of Ref.~\cite{Godazgar2010},
\begin{equation}
\epsilon(F)=\frac{\ii}{8}\phi(F),\qquad
F_{ab}=\ii\,\tr\!\left(\Gamma_{ab}\CC^{-1}\epsilon(F)\right).
\label{eq:sinverse}
\end{equation}

The Lorentzian real structure is also fixed by these matrices.
With \(\mathsf R=\gamma_5\), one has
\begin{equation}
\mathsf R^*\mathsf R=-I,\qquad
\mathsf R^T\CC\mathsf R=-\CC,\qquad
\phi(F)=-\mathsf R^T\phi(F)^*\mathsf R,\qquad
\epsilon(F)=\mathsf R^T\epsilon(F)^*\mathsf R
\label{eq:sreality}
\end{equation}
for real \(F\). The Weyl spinor obeys the corresponding quartic reality
condition. These conditions constrain complex factors; they do not
replace factorization over \(\mathbb C\) by real factorization.
This is the De Smet convention
\cite{DeSmet2002,DeSmet2003,Godazgar2010}.

\subsection{Rank-one Clifford proof}

For arbitrary commuting \(\zeta\), define
\begin{equation}
X^{ab}=\zeta^T\CC\Gamma^{ab}\zeta,\qquad
N=\zeta\zeta^T\CC,\qquad
\Omega=X_{ab}\Gamma^{ab}.
\label{eq:sX}
\end{equation}
The sixteen matrices \(I,\Gamma^a,\Gamma^{ab}\) form a basis of
\(\operatorname{End}(\mathbb C^4)\). Their traces give
\begin{equation}
\begin{gathered}
\tr(\Gamma_a\Gamma_b)=4\eta_{ab},\qquad
\tr(\Gamma_{ab}\Gamma_{cd})
=-4(\eta_{ac}\eta_{bd}-\eta_{ad}\eta_{bc}),\\
\mathsf Z=\frac14\tr(\mathsf Z)I
+\frac14\tr(\Gamma_a\mathsf Z)\Gamma^a
-\frac18\tr(\Gamma_{ab}\mathsf Z)\Gamma^{ab}
\end{gathered}
\label{eq:scompleteness}
\end{equation}
for any \(4\times4\) matrix \(\mathsf Z\).
Since \(\CC\) and \(\CC\Gamma_a\) are antisymmetric,
\(\tr N=\tr(\Gamma_aN)=0\). Hence
\begin{equation}
N=-\Omega/8,\qquad N^2=0,\qquad N\Gamma^cN=0 .
\label{eq:snilpotent}
\end{equation}
Clifford multiplication yields
\begin{equation}
\Omega^2=-2X_{ab}X^{ab}I+X_{ab}X_{cd}\Gamma^{abcd}.
\label{eq:somega}
\end{equation}
The scalar and four-gamma terms are independent in five dimensions.
Their vanishing implies \(X_{ab}X^{ab}=0\) and
\(X^{[ab}X^{cd]}=0\). To obtain the stronger metric-contraction
identity, use
\begin{equation}
[\Omega,\Gamma^c]=4X_a{}^c\Gamma^a,\qquad
\tr(\Gamma^d\Gamma^a\Omega)=-8X^{da}.
\label{eq:scommutator}
\end{equation}
Taking a trace of
\(0=\Omega\Gamma^c\Omega=\Gamma^c\Omega^2+
4X_a{}^c\Gamma^a\Omega\) with \(\Gamma^d\) gives
\begin{equation}
X^{[ab}X^{cd]}=0,\qquad X^{da}X_a{}^c=0 .
\label{eq:snullidentities}
\end{equation}

Every trace-subtraction term in
\(\W_g(F)_{abcd}X^{ab}X^{cd}\) vanishes by the second identity.
Writing \(q=F_{ab}X^{ab}\), the remaining contraction is
\begin{equation}
T(F)_{abcd}X^{ab}X^{cd}
=q^2+F_{ac}F_{bd}X^{ab}X^{cd}.
\label{eq:sTcontraction}
\end{equation}
The first identity in Eq.~\eqref{eq:snullidentities}, expanded as
\(X^{ab}X^{cd}-X^{ac}X^{bd}+X^{ad}X^{bc}=0\) and contracted with
\(F_{ab}F_{cd}\), gives
\(F_{ac}F_{bd}X^{ab}X^{cd}=q^2/2\).
Therefore, for every complex two-form,
\begin{equation}
P_{\W_g(F)}=\frac32q_F^2,\qquad
\Psi[\W_g(F)]_{ABCD}=\frac32\phi(F)_{(AB}\phi(F)_{CD)}.
\label{eq:ssquare}
\end{equation}
In the normalized convention \eqref{eq:sinverse}, this is
\begin{equation}
P_{\W_g(F)}=-96\,[\epsilon(F)_{AB}\zeta^A\zeta^B]^2 .
\label{eq:snormalizedsquare}
\end{equation}
The coefficient follows from \((3/2)(8/\ii)^2=-96\).
This proof is pointwise and independent of the black-ring curvature.

\section{Irreducibility and De Smet type 22}
\label{sec:rank}

By Eqs.~\eqref{eq:scopy} and \eqref{eq:ssquare},
\begin{equation}
\begin{gathered}
P_C=\frac32\left(\frac{q_E^2}{S_e}+\frac{q_H^2}{S_H}\right)
=\frac32(Q_H-Q_e)(Q_H+Q_e),\\
Q_e=\frac{q_E}{\sqrt{-S_e}},\qquad
Q_H=\frac{q_H}{\sqrt{S_H}}=\frac{q_M}{\sqrt{S_m}} .
\end{gathered}
\label{eq:sfactors}
\end{equation}
Both scalar square roots are positive on the open exterior.
Other local branch choices exchange or rescale the factors and do not
affect their irreducible degrees or multiplicities.
Since \(E_{yt}=-\lambda/F_x\neq0\) and \(M\) has only an
\(x\bar\phi\) component, \(E\) and \(M\) are independent.
Invertibility of Eq.~\eqref{eq:sspinmaps} proves that the two factors
are nonzero and nonproportional.

For their ranks, use the equivalent quadratics
\begin{equation}
f_\pm=q_E\pm c q_M,\qquad
c=R\sqrt{\frac{\lambda\nu}{F_x}},\qquad
P_C=\frac{3}{2S_e}f_-f_+.
\label{eq:srawfactors}
\end{equation}
They differ from the factors in Eq.~\eqref{eq:sfactors} only by
nonzero scalar multiples and an interchange.
On the exterior patch \(F_y>0\), take the orthonormal coframe
\begin{equation}
\begin{aligned}
e^0&=\sqrt{\frac{F_y}{F_x}}
\left(\dd t-C_0R\frac{1+y}{F_y}\dd\bar\psi\right),&
e^1&=\frac R\Delta\sqrt{\frac{F_x}{G_x}}\,\dd x,&
e^2&=\frac R\Delta\sqrt{-\frac{F_x}{G_y}}\,\dd y,\\
e^3&=\frac R\Delta\sqrt{G_x}\,\dd\bar\phi,&
e^4&=\frac R\Delta\sqrt{-\frac{F_xG_y}{F_y}}\,\dd\bar\psi .
\end{aligned}
\label{eq:sframe}
\end{equation}
Substituting Eq.~\eqref{eq:sEcomponents} gives
\begin{equation}
\begin{gathered}
E=a\,e^0\wedge e^1+b\,e^0\wedge e^2+d\,e^2\wedge e^4,
\qquad cM=h\,e^1\wedge e^3,\\
a=-\frac{\lambda\Delta\sqrt{F_yG_x}}{RF_x^2},\qquad
b=\frac{\lambda\Delta\sqrt{-G_y}}{RF_x\sqrt{F_y}},\qquad
d=\frac{C_0(1-\lambda)\Delta^2}{RF_x^2\sqrt{F_y}},\qquad
h=\frac{\sqrt{\lambda\nu}\Delta^2}{RF_x}.
\end{gathered}
\label{eq:sframefields}
\end{equation}
The \(e^1\wedge e^4\) coefficient cancels.
With Eq.~\eqref{eq:sgamma}, the symmetric matrices of \(f_\pm\) are
\begin{equation}
\mathsf B(\pm h)=
2\begin{pmatrix}
\mp h&\ii d&a&-b\\
\ii d&\pm h&-b&-a\\
a&-b&\mp h&-\ii d\\
-b&-a&-\ii d&\pm h
\end{pmatrix}.
\label{eq:sBmatrix}
\end{equation}
Their determinants are equal:
\begin{equation}
\begin{split}
\det\mathsf B(\pm h)
&=16\left[(a^2+b^2+d^2-h^2)^2+4b^2(h^2-d^2)\right]\\
&=\frac{16\lambda^2\Delta^6}{R^4F_x^6}D(x,y),
\end{split}
\label{eq:sdet}
\end{equation}
where
\begin{equation}
\begin{split}
D={}&4\nu^2F_x^2y^2
+4\nu(\lambda-\nu)(\lambda+2x+\lambda x^2)y\\
&+\nu^2[\lambda(1-x^2)-2x]^2+4\lambda(\lambda-2\nu).
\end{split}
\label{eq:sD}
\end{equation}
This calculation is a four-by-four determinant followed by substitution
of the four coefficients in Eq.~\eqref{eq:sframefields}.

\subsection{Strict positivity throughout the open exterior}

Regard \(D\) as a quadratic in \(y\), whose leading coefficient is
\(4\nu^2F_x^2>0\). Its discriminant is
\begin{equation}
\operatorname{disc}_yD
=16\lambda\nu^2(1-x^2)(2-\lambda+\lambda x)
(2+\lambda+\lambda x)(1+\nu x)
[\nu(2+\lambda x)-\lambda].
\label{eq:sdiscriminant}
\end{equation}
Every factor except the last is strictly positive in the prescribed
parameter and \(x\) ranges. A negative discriminant makes \(D\)
positive for all real \(y\).
Otherwise \(\nu\geq\nu_c=\lambda/(2+\lambda x)\).
Writing \(L=\lambda+2x+\lambda x^2\), the vertex is
\begin{equation}
y_0=-\frac{(\lambda-\nu)L}{2\nu F_x^2},\qquad
y_0(\nu_c)+1=\frac{(1-x)(2-\lambda+\lambda x)}{2F_x}>0.
\label{eq:svertex}
\end{equation}
If \(L\leq0\), then \(y_0\geq0\).
If \(L>0\), then \(y_0\) increases with \(\nu\), so the second
formula implies \(y_0>-1\).
Finally,
\begin{equation}
D(x,-1)=J(x)^2,\qquad
J(x)=-2\lambda+2\nu+\lambda\nu+2\nu x+\lambda\nu x^2 .
\label{eq:sJ}
\end{equation}
The convex quadratic is strictly decreasing for \(y<-1\), giving
\(D(x,y)>D(x,-1)\geq0\). Thus \(D>0\) for all
\(-1<x<1\), \(y<-1\), and every fixed admissible pair
\((\lambda,\nu)\).

A nonzero complex quadratic is reducible precisely when its symmetric
matrix has rank at most two: the matrix of a product of two linear
forms has rank at most two, and a rank-one or rank-two form can be
factored over \(\mathbb C\). Equations~\eqref{eq:sdet}--\eqref{eq:sJ}
therefore prove that both quadratics are irreducible.
Together with their distinctness, this establishes De Smet type \(22\)
at every open-exterior point of every admissible ring, including the
balanced and static families.

For comparison, a shorter argument suffices for generic irreducibility:
\(D(-1,-1)=4\lambda^2(1-\nu)^2>0\). For each fixed parameter pair,
\(D(x,y)\) is therefore not the zero polynomial; it is nonzero on an
open dense set. The strict-positivity argument above removes any
remaining interior exceptional locus.

\subsection{Ergosurface and horizon}

Under a change of oriented spin frame, a quadratic matrix transforms
by \(\mathsf B\mapsto S^T\mathsf BS\).
The spin generators are traceless, so \(\det S=1\), also in the
connected complexified spin group. Hence its determinant is invariant.
Although the stationary coframe \eqref{eq:sframe} fails at \(F_y=0\),
Eq.~\eqref{eq:sdet} has no \(F_y\) denominator and extends to a regular
spin frame. We note there is no De Smet type change at the ergosurface.

On the future horizon,
\begin{equation}
\left.\det\mathsf B(\pm h)\right|_{\mathcal H^+}
=\frac{16\lambda^2(1+\nu x)^6}{\nu^6R^4F_x^6}
[2-\lambda\nu+2\nu x+\lambda\nu x^2]^2>0.
\label{eq:shorizondet}
\end{equation}
The bracket has derivative \(2\nu F_x>0\) and its value at \(x=-1\)
is \(2(1-\nu)>0\).
Independence of the fields continues across the horizon:
\begin{equation}
\xi^aE_{ab}=-\partial_b(\xi^2)
=\partial_b(F_y/F_x),\qquad
\xi^aH_{ab}=0 ,
\label{eq:shorizonindependence}
\end{equation}
and the \(y\) derivative of \(F_y/F_x\) is \(\lambda/F_x\neq0\).
Regularity of the underlying fields is established in
Sec.~\ref{sec:regularity}.
Thus the regular future horizon, away from the axes, is also type \(22\).

\section{Static family, axes, and parameter endpoints}
\label{sec:exceptional}

For \(\lambda=\nu>0\), \(C_0=d=0\).
The open static exterior has \(F_y>0\), \(b^2>0\), and \(h^2>0\), so
\begin{equation}
\det\mathsf B(\pm h)
=16[(a^2+b^2-h^2)^2+4b^2h^2]>0 .
\label{eq:sstaticdet}
\end{equation}
Both Maxwell contributions survive, and the factors remain distinct.
The nonzero static ring is therefore type \(22\) on its open exterior
and regular open horizon. At
\((\nu,\lambda,R,x,y)=(1/3,1/3,1,0,-3/2)\), the determinants of
\(q_E\pm q_M/3\) are \(23/2\).

\subsection{Regular \texorpdfstring{\(\psi\)}{psi} axis}

At \(y=-1\), \(-1<x<1\), Eq.~\eqref{eq:sdet} becomes
\begin{equation}
\det\mathsf B(\pm h)
=\frac{16\lambda^2(1+x)^6}{R^4F_x^6}J(x)^2 .
\label{eq:spsiaxis}
\end{equation}
Here \(J'(x)=2\nu F_x>0\),
\(J(-1)=-2\lambda(1-\nu)<0\), and
\(J(1)=2[2\nu-\lambda(1-\nu)]\).
There is a unique zero \(x_*\in(-1,1)\) precisely when
\begin{equation}
\lambda<\frac{2\nu}{1-\nu},\qquad J(x_*)=0 .
\label{eq:saxiscondition}
\end{equation}
This includes every balanced ring and every nonzero static ring.
Away from \(x_*\), this regular axis is type \(22\).

To determine the type at \(x_*\), evaluate the limiting quadratic in
a smooth axis frame. In Eq.~\eqref{eq:sframefields}, \(b=0\),
\(a\neq0\), and \(h^2=a^2+d^2\).
For \(\mathsf B(h)/2\), use the two-by-two blocks
\begin{equation}
K=\begin{pmatrix}-h&\ii d\\\ii d&h\end{pmatrix},\qquad
L=\operatorname{diag}(a,-a),\qquad
K'=\begin{pmatrix}-h&-\ii d\\-\ii d&h\end{pmatrix}.
\label{eq:saxisblocks}
\end{equation}
Then
\begin{equation}
\det K=d^2-h^2=-a^2\neq0,\qquad
K'-LK^{-1}L
=\frac{a^2+d^2-h^2}{d^2-h^2}K'=0.
\label{eq:saxisschur}
\end{equation}
Each quadratic has rank exactly two and hence splits into two distinct
linear factors. The two quadratics cannot share a linear factor:
their difference is \(2cq_M\), whose matrix has rank four.
The Weyl polynomial therefore has four distinct linear factors at
this smooth-axis locus, giving type \(1111\).
This exception is compatible with the generic type-\(22\) classification.

\subsection{\texorpdfstring{\(\phi\)}{phi} axes}

The limiting determinants on the outer and inner \(\phi\) axes are
\begin{equation}
\begin{aligned}
\left.\det\mathsf B(\pm h)\right|_{x=-1}
&=\frac{64\lambda^2(1+y)^6}{(1-\lambda)^6R^4}
[\lambda-\nu-\nu y+\lambda\nu y]^2,\\
\left.\det\mathsf B(\pm h)\right|_{x=1}
&=\frac{64\lambda^2(1-y)^6}{(1+\lambda)^6R^4}
[\lambda-\nu+\nu(1+\lambda)y]^2.
\end{aligned}
\label{eq:sphiaxes}
\end{equation}
For finite exterior \(y<-1\), the first bracket is greater than
\(\lambda(1-\nu)>0\). On the inner axis an exterior zero occurs only
when \(\lambda>2\nu/(1-\nu)\), at
\(y_*=-(\lambda-\nu)/[\nu(1+\lambda)]\).
That regime is unbalanced and the inner axis is conically singular.
The expression gives the limit of the smooth-region polynomial, not
a smooth vacuum classification of the distributional defect.
For balanced rings both axes are regular and have no such finite
exterior determinant zero.
Axis intersections and the parameter-threshold corner are not included
in the open-axis statements.

The flat endpoint \(\lambda=\nu=0\) has vanishing Weyl curvature and type
O. Expressions containing \(1/\lambda\) or \(1/\nu\) are not evaluated
there by substitution. The endpoints \(\lambda=1\), \(\nu=1\),
curvature singularities, and separately scaled limiting geometries
are outside the finite-ring theorem.

\section{Flux, potentials, and regularity}
\label{sec:regularity}

With \(\phi\) of period \(2\pi\) and the orientation of
\(\dd x\wedge\dd\phi\), the linking sphere satisfies
\begin{equation}
\int_{S^2}H
=\int_{-1}^{1}\dd x\int_0^{2\pi}\dd\phi=4\pi .
\label{eq:sflux}
\end{equation}
The two local potentials
\begin{equation}
A_+=(x-1)\dd\phi,\qquad
A_-=(x+1)\dd\phi,\qquad
A_+-A_-=-2\dd\phi
\label{eq:spotentials}
\end{equation}
cover the respective polar caps and satisfy \(\dd A_\pm=H\).
On a simply connected part of their overlap the difference is
\(\dd(-2\phi)\).
Near a regular \(\phi\) axis, \(1\mp x\) is proportional to
\(\rho^2\). Thus \(A_\pm\) is proportional to
\(\rho^2\dd\phi=X\dd Y-Y\dd X\) near its own cap, and
\(H\) is proportional to
\(\rho\,\dd\rho\wedge\dd\phi=\dd X\wedge\dd Y\).
The field is smooth at the axes although the angular coordinate is not.
On a regular \(\psi\) axis the coordinates \(x,\phi\) do not collapse.

The nonzero integral \eqref{eq:sflux} excludes a globally smooth
potential on the linking sphere. It also excludes expressing \(H\)
globally as a constant linear combination of exact Killing two-forms.
The local algebraic obstruction for the full curvature is the
independent statement proved in Sec.~\ref{sec:obstruction}.
Within the restricted ansatz \(F=f(x,y)\dd x\wedge\dd\phi\), closure
forces \(\partial_y f=0\), and the Maxwell divergence then forces
\(\partial_x f=0\). This selects the magnetic representative up to a
constant within that ansatz, without implying uniqueness among all
Maxwell solutions.

The balanced metric has a regular future-horizon extension.
In an ingoing chart for the rotating ring, the shifts of \(t\) and
\(\bar\psi\) depend only on \(y\); \(x\) and \(\phi\) are unchanged
\cite{EmparanReall2006}. Hence \(H\) extends smoothly.
The stationary Killing vector and the metric are smooth in that
extension, so \(E=\dd\xi^\flat\) is smooth as well.
The static family can be treated in its corresponding regular
time-radial chart. On the open horizon, the coefficients are
\begin{equation}
\left.S_e\right|_{\mathcal H^+}
=-\frac{\lambda(x+1/\nu)}{F_x},\qquad
\left.S_H\right|_{\mathcal H^+}
=\frac{x+1/\nu}{R^2\nu\alpha^2},
\label{eq:shorizonscalars}
\end{equation}
which are finite and nonzero.
The same fields are smooth through the stationary limit.
Consequently the curvature identity extends across the regular
future horizon and ergosurface. For unbalanced metrics, regularity
claims apply only where the metric itself is smooth.

Finally, the quadratic factors in Eq.~\eqref{eq:sfactors} should be
distinguished from additional Maxwell solutions. For scalar weights
\(u(x,y),v(x,y)\),
\begin{equation}
\begin{aligned}
\dd(uE+vH)&=\dd u\wedge E+\dd v\wedge H,\\
\nabla_a(uE^{ab}+vH^{ab})
&=(\partial_a u)E^{ab}+(\partial_a v)H^{ab} \,,
\end{aligned}
\label{eq:sweightedfields}
\end{equation}
and the original Maxwell equations do not set these extra terms to zero.

\section{Comparison with reported Weyl polynomial}
\label{sec:comparison}

Section 3.3 of Ref.~\cite{Godazgar2010} states that the rotating-ring
polynomial is generically irreducible and remains irreducible in the
static limit. The proof in Secs.~\ref{sec:tensor}--\ref{sec:rank} gives instead two distinct irreducible quadratics.
This conclusion uses the metric, the quadratic map, and the Clifford
correspondence and does not use the Appendix~E polynomial as a premise.

For a direct comparison, at Eq.~\eqref{eq:switness} we use the coframe
\eqref{eq:sframe} and spinor argument \(\zeta=(u,v,w,z)\). The
polynomials computed from the two-forms are
\begin{equation}
\begin{aligned}
q_E={}&\frac{72\ii}{5}\sqrt{\frac25}(uv-wz)
-\frac95\sqrt{\frac25}(uw-vz)-9(vw+uz),\\
q_M={}&-\frac92(u^2-v^2+w^2-z^2),\\
P_C={}&\frac13(q_M-\sqrt5\,q_E)(q_M+\sqrt5\,q_E).
\end{aligned}
\label{eq:spointpolynomial}
\end{equation}
Contraction of the directly computed Weyl tensor with the gamma
matrices gives the same quartic in all 35 coefficient slots.
The factors \(q_M\pm\sqrt5\,q_E\) both have determinant \(45927/50\),
and their coefficient vectors are independent.
For the factors \(f_\pm\) of Eq.~\eqref{eq:srawfactors}, the determinant
at this point is instead \(45927/1250\); multiplication of a quadratic
matrix by \(\sqrt5\) multiplies its determinant by \(25\).

The literal source of Appendix~E has two independent problems.
First, its coefficient \(A_6\) includes
\begin{equation}
v^2F_x^2\left[
\Delta^2\nu+\lambda(1-y^2)+y\lambda\nu(1-xy)
+xy\lambda\nu\Delta\right],
\label{eq:stypo}
\end{equation}
where \(v\) is a spinor variable, whereas \(\nu\) is a ring parameter.
Since \(A_6\) multiplies \(u^2z^2+v^2w^2\), this term generates
degree-six monomials. At Eq.~\eqref{eq:switness}, their contribution
to the printed polynomial is \(-54v^4w^2-54u^2v^2z^2\).
A Weyl polynomial defined by a rank-four spinor is homogeneous of
degree four. Replacing \(v^2\) by \(\nu^2\) repairs this degree mismatch
but fails to make the full expression correct.

Second, the printed coefficient
\begin{equation}
A_3=4F_xF_y\sqrt{-G_xG_y}
\label{eq:sA3}
\end{equation}
fails the regular flat limit. At \(\lambda=\nu=0\), the coordinate
transformation
\begin{equation}
\rho=\frac{R\sqrt{1-x^2}}{x-y},\qquad
\zeta_{\rm b}=\frac{R\sqrt{y^2-1}}{x-y}
\label{eq:sflatcoordinates}
\end{equation}
gives
\begin{equation}
\dd s^2=-\dd t^2+\dd\rho^2+\rho^2\dd\bar\phi^2
+\dd\zeta_{\rm b}^2+\zeta_{\rm b}^2\dd\bar\psi^2.
\label{eq:sflatmetric}
\end{equation}
The mixed terms cancel using
\((xy-1)^2+(1-x^2)(y^2-1)=(x-y)^2\).
The Weyl polynomial must vanish. All other printed \(A_i\) vanish
in this limit, but Eq.~\eqref{eq:sA3} leaves
\begin{equation}
P_{\rm printed}
=\frac{24\ii(x-y)}{R^2}\sqrt{(1-x^2)(y^2-1)}
(u^2-v^2+w^2-z^2)(uw-vz),
\label{eq:sflatcontradiction}
\end{equation}
which is nonzero at ordinary points with \(-1<x<1\), \(y<-1\).
For example, at \(R=1,x=0,y=-2\) and
\((u,v,w,z)=(1,0,1,0)\), it equals \(96\ii\sqrt3\).
Neither a nonzero overall normalization nor an invertible spinor-basis
change can turn a nonzero polynomial into the zero flat-space polynomial.

The classification correction concerns De Smet factorization only.
Its relation to the CMPP and spinor-helicity descriptions is discussed
in Refs.~\cite{Godazgar2010,MNO2019}; this does not imply a change in
the ring's null-alignment classification.

\section{Little-group structure and real null alignment}
\label{sec:mno}

The little-group decomposition of Ref.~\cite{MNO2019} resolves the
relation between the complex factorization in Sec.~\ref{sec:rank}
and real null alignment. The Weyl identity and the irreducibility
theorem remain the inputs to this analysis.

\subsection{Conventions and weighted factors}

We use a real null frame \((k,n,m_1,m_2,m_3)\) with
\(k\cdot n=-1\), \(m_i\cdot m_j=\delta_{ij}\), and all other
inner products zero. Tensor labels \(0,1\) denote \(k,n\), while
\(i,j=1,2,3\) label screen vectors; for example,
\(C_{0i1j}=C(k,m_i,n,m_j)\).
The five Weyl spinor blocks \(\Psi^{(r)}\), \(r=0,\ldots,4\),
have boost weights \(2-r\) and dimensions \(5,8,9,8,5\).
Their fine irreducible pieces are five real symmetric trace-free
matrices \(\psi^{(r)}_{ij}\), three real vectors
\(\chi^{(1,2,3)}_i\), and the scalar
\(T=\Psi_{\rm tr}^{(2)}\), accounting for all 35 Weyl components.

For an explicit tensor dictionary, let
\(\varepsilon_{123}=+1\) be the numerical alternating symbol and
\(\operatorname{STF}M=(M+M^T)/2-\operatorname{tr}(M)I/3\).
The oriented spinor convention used here corresponds to
\(\epsilon^{\rm MNO}_{ijk}=-\varepsilon_{ijk}\).
With these signs the dictionary is
\begin{equation}
\begin{gathered}
M_{0,i\ell}=\varepsilon_{jk\ell}C_{0ijk},\qquad
M_{1,i\ell}=\varepsilon_{jk\ell}C_{1ijk},\\
\begin{aligned}
\psi^{(0)}_{ij}&=8C_{0i0j},&
\psi^{(4)}_{ij}&=8C_{1i1j},\\
\psi^{(1)}&=-2\sqrt2\,\operatorname{STF}M_0,&
\chi^{(1)}_i&=-8\sqrt2\,C_{010i},\\
\psi^{(3)}&=2\sqrt2\,\operatorname{STF}M_1,&
\chi^{(3)}_i&=8\sqrt2\,C_{011i},\\
T&=16C_{0101},&
\chi^{(2)}_i&=4\varepsilon_{ijk}C_{01jk},\\
\psi^{(2)}_{ij}&=-4(C_{0i1j}+C_{0j1i})-\frac{T}{6}\delta_{ij}.
\end{aligned}
\end{gathered}
\label{eq:smnodictionary}
\end{equation}
This fixes the signs and normalizations of all entries below.
It is the real tensor form of the spinor projections in
Eqs.~(5.6)--(5.9) of Ref.~\cite{MNO2019}.

Set
\begin{equation}
\begin{gathered}
A=\frac{E}{\sqrt{-S_e}},\qquad B=\frac{H}{\sqrt{S_H}},\qquad
\mathcal U=A+B,\qquad \mathcal V=B-A,\\
C_E=\frac{\W(E)}{S_e}=-\W(A),\qquad
C_H=\frac{\W(H)}{S_H}=\W(B),\qquad
P_C=\frac32q_{\mathcal U}q_{\mathcal V}.
\end{gathered}
\label{eq:smnoroots}
\end{equation}
The last equality is Eq.~\eqref{eq:sfactors} in the same
quadratic normalization. The calligraphic notation distinguishes
these factors from the simple string bivectors \(U,V\) below.
They are position-dependent weighted two-forms; the differential
terms in Eq.~\eqref{eq:sweightedfields} still apply.

\subsection{Real WAND criterion and exterior regions}

For any real null \(k\), define screen vectors
\(a_i=A(k,m_i)\), \(b_i=B(k,m_i)\).
Contracting the fixed Weyl-square map gives
\begin{equation}
C_{0i0j}=\frac32\operatorname{STF}(b_i b_j-a_i a_j),
\qquad
C_{0i0j}=0\quad\Longleftrightarrow\quad b=\pm a .
\label{eq:smnowand}
\end{equation}
Indeed, a vanishing trace-free part requires
\(bb^T-aa^T=cI_3\). The left side has rank at most two, so
\(c=0\); over the real Euclidean screen this gives \(b=\pm a\).
Vanishing screen contraction means that contraction of
\(\mathcal U\) or \(\mathcal V\) with \(k\) is proportional
to \(k^\flat\). Thus a real null direction is a WAND precisely
when it is an eigenvector of \(g^{-1}\mathcal U\) or
\(g^{-1}\mathcal V\). This criterion covers every real null
direction and follows from the established tensor identity.

The exact component calculations use the balanced parameters and
three exterior points
\begin{equation}
\begin{gathered}
R=1,\qquad \nu=\frac13,\qquad\lambda=\frac35,\qquad
\alpha=\frac{3}{\sqrt{10}},\\
p_a:\ (x,y)=\left(-\frac12,-\frac43\right),\qquad
p_b:\ (x,y)=\left(-\frac13,-\frac32\right),\qquad
p_G:\ (x,y)=\left(0,-\frac65\right).
\end{gathered}
\label{eq:smnopoints}
\end{equation}
All use the physical angles of Eq.~\eqref{eq:sangles} and lie
away from axes, the stationary limit and the horizon.
At \(p_a,p_b\), each weighted factor has a real null eigenpair.
The reflection \(P:\phi\mapsto-\phi\) exchanges the factors
up to sign and supplies four distinct real WANDs in total.
The nonzero weight-one blocks show that all four are simple:
these points are type \(\mathrm{I}_i\).

At \(p_G\), use the proportional representatives
\(F_\pm=E\pm\sqrt{-S_e/S_H}\,H\) and \(L_\pm=g^{-1}F_\pm\).
For \(L_+\), the exact verification is
\begin{equation}
\begin{gathered}
\det(zI-L_+)=z\left(z^4+\frac{8424}{15625}z^2
+\frac{419904}{244140625}\right),\\
\ker L_+=\operatorname{span}\{K\},\qquad
K=\left(\frac{2\sqrt{10}}5,0,0,-\frac{8\sqrt5}{25},1\right),
\qquad g(K,K)=-\frac1{200}.
\end{gathered}
\label{eq:smnoGcertificate}
\end{equation}
The components of \(K\) are in the coordinate order
\((t,x,y,\phi,\psi)\).
The quadratic in \(z^2\) has positive coefficients and
discriminant \(13856832/48828125>0\), so both its roots are
negative. The only real eigenline is therefore the timelike
kernel and the reflection gives the same conclusion for \(L_-\).
Neither factor has a real null eigenvector: \(p_G\) is type
\(\mathrm G\), although the complex De Smet factors remain.

The strict spectral signs, distinct WANDs and nonzero weight-one
tests persist in neighborhoods of the respective samples.
These establish open exterior regions of types
\(\mathrm G\) and \(\mathrm{I}_i\); the sample verifications
do not determine their complete boundaries.
The horizon is type \(\mathrm{II}\)
as noted in \cite{PravdaPravdova2005}. Under the convention assigning a
spacetime its most general pointwise type, the ring is
\(\mathrm G\) \cite{Godazgar2010}. The earlier real-WAND branch analysis in
Ref.~\cite{PravdaPravdova2005} leaves open alternative branches
where its displayed branch becomes nonreal;
Eq.~\eqref{eq:smnowand} provides the exhaustive criterion used
at \(p_G\). None of these real-alignment statements changes the
complex factorization or its horizon and axis qualifications.

\subsection{Relations among the fine irreducible components}

At both \(p_a\) and \(p_b\), choose \(k,n\) either from the same
weighted factor or as a reflection-related pair from different
factors. The exact same-frame decompositions of \(C_E,C_H,C\)
give Table~\ref{tab:mnosupport}.

\begin{table}[ht]
\centering
\begin{tabular}{lccc|ccc}
\hline\hline
&\multicolumn{3}{c|}{Same factor}
&\multicolumn{3}{c}{Reflection pair, different factors}\\
Irrep & \(C_E\)&\(C_H\)&\(C\)&\(C_E\)&\(C_H\)&\(C\)\\
\hline
\(\psi^{(0)},\psi^{(4)}\)&NZ&NZ&0&NZ&NZ&0\\
\(\psi^{(1)},\psi^{(3)}\)&NZ&NZ&NZ&NZ&0&NZ\\
\(\psi^{(2)},\chi^{(1)},\Psi_{\rm tr}^{(2)},\chi^{(3)}\)
&NZ&NZ&NZ&NZ&NZ&NZ\\
\(\chi^{(2)}\)&NZ&NZ&NZ&0&0&0\\
\hline\hline
\end{tabular}
\caption{Fine-irrep support at each of the two exact points
\(p_a,p_b\), in the specified paired-WAND frames.
NZ means that each listed complete irrep is nonzero.
The outer spin-2 blocks cancel between nonzero contributions;
the cross-factor \(\chi^{(2)}\) zero holds separately for
each contribution.}
\label{tab:mnosupport}
\end{table}

In both choices of paired-WAND frame,
\begin{equation}
\psi^{(r)}[C_E]=-\psi^{(r)}[C_H]\neq0,\qquad r=0,4 .
\label{eq:smnocancel}
\end{equation}
In the same-factor frames used here, \(\mathcal U\) has no leading or
trailing Maxwell block. Its quadratic is purely of bidegree
\((1,1)\) in the two \(k\)-spinor and two \(n\)-spinor variables.
Multiplication by \(q_{\mathcal V}\) leaves Weyl bidegrees
\((3,1),(2,2),(1,3)\), or boost weights \(+1,0,-1\).
Expanding this product and applying the linear projections
\eqref{eq:smnodictionary} fixes the relations among the surviving
irreps. The complete arrays obey
\(\mathcal I[C]=\mathcal I[C_E]+\mathcal I[C_H]\)
for every fine irrep \(\mathcal I\).

For the reflection pair, \(P\) exchanges \(k,n\) and fixes the
screen. The field \(E\) is \(P\)-even and \(H\) is \(P\)-odd,
so both quadratic contributions are even.
Consequently \(C_{01ij}=0\) in each contribution separately,
and hence \(\chi^{(2)}[C_E]=\chi^{(2)}[C_H]=0\).
This is a parity zero, distinct from the cancellations in
Eq.~\eqref{eq:smnocancel}.

At each of \(p_a,p_b\), the minimum over all real frames whose
first null direction is a WAND, allowing an arbitrary second
null direction, is six nonzero fine irreps. Reflection pairs
attain this minimum with
\begin{equation}
\begin{gathered}
\psi^{(1)},\ \chi^{(1)},\ \psi^{(2)},\
\Psi_{\rm tr}^{(2)},\ \psi^{(3)},\ \chi^{(3)}\neq0,\\
\psi^{(0)}=\chi^{(2)}=\psi^{(4)}=0 .
\end{gathered}
\label{eq:smnominimum}
\end{equation}
For a fixed WAND, \(\psi^{(1)}\) and \(\chi^{(1)}\) remain
nonzero under null rotations. Fewer than six occupied irreps
would require three of the remaining six to vanish.
At each sample, all 20 corresponding component ideals have
reduced Gr\"obner basis \([1]\), even over the complexified
three-parameter null-rotation space. Exact discrete isometries
cover all four WANDs; screen rotations and boosts preserve
vanishing. This proves the stated lower bound with unrestricted
second null direction. It is a minimum at these two points,
not a theorem throughout the entire \(\mathrm{I}_i\) region;
the displayed support belongs to the specified reflection pairs.

\subsection{Comparison}

In the Myers--Perry controls using the principal frames of
Ref.~\cite{PravdaEtAl2007}, only the zero-weight
pieces \(\psi^{(2)},\chi^{(2)},\Psi_{\rm tr}^{(2)}\) survive.
The ring samples require weight \(+1\) and \(-1\) data as well.
In the normalized product frame of the string limit below, with
\(\hat z_i\) the unit screen vector along the string,
\begin{equation}
\psi^{(2)}_{ij}=-\frac{4r_0}{r^3}
\left(\frac{\delta_{ij}}3-\hat z_i\hat z_j\right),\qquad
\Psi_{\rm tr}^{(2)}=-\frac{16r_0}{r^3},\qquad
\chi^{(2)}=0 .
\label{eq:smnostring}
\end{equation}
These are the string components of Ref.~\cite{MNO2019}.
The relation
\(\psi^{(2)}_{zz}+\Psi_{\rm tr}^{(2)}/6=0\)
expresses the cancellation of string-direction curvature.
Both nonzero final irreps survive; the four zero bivector modes arise without eliminating a complete
additional irrep.

The component dictionary reproduces the metric-derived
Tangherlini and Schwarzschild-string controls at two exact
regular points each and passes an oriented conversion check on
a set spanning all 35 Weyl components.
The finite-ring component calculations use exterior frames.
The horizon statements here rely on the regularity proof in
Sec.~\ref{sec:regularity} and the null-alignment result cited above.

\section{Black-string limit}
\label{sec:string}

Taking the balanced scaling
\begin{equation}
\nu=\frac{r_0}{R},\qquad
\lambda=\frac{2\nu}{1+\nu^2},\qquad
y=-\frac Rr,\qquad x=\cos\theta,\qquad z=R\psi,\qquad R\to\infty .
\label{eq:slimit}
\end{equation}
Then \(\alpha\to1\), \(C_0R\to\sqrt2r_0\).
The \(t,z\) part of the limiting metric has
\(g_{tt}=-1+2r_0/r\), \(g_{tz}=-\sqrt2r_0/r\), and
\(g_{zz}=1+r_0/r\). The local boost
\begin{equation}
T=\sqrt2\,t-z,\qquad Z=-t+\sqrt2\,z
\label{eq:sboost}
\end{equation}
puts the metric in the product form
\begin{equation}
\dd s^2=-f(r)\dd T^2+\frac{\dd r^2}{f(r)}
+r^2(\dd\theta^2+\sin^2\theta\,\dd\phi^2)+\dd Z^2,
\qquad f(r)=1-\frac{r_0}{r}.
\label{eq:sstringmetric}
\end{equation}
This is just the Schwarzschild string of Ref.~\cite{EmparanReall2006}.
The finite-ring stationary vector tends to
\(\sqrt2\,\partial_T-\partial_Z\).
With
\(e^0=\sqrt f\,\dd T\), \(e^1=\dd r/\sqrt f\),
\(e^2=r\dd\theta\), \(e^3=r\sin\theta\,\dd\phi\),
and \(e^4=\dd Z\), define \(U=e^0\wedge e^1\),
\(V=e^2\wedge e^3\).
The limiting fields and scalar partners are
\begin{equation}
E\longrightarrow\frac{\sqrt2r_0}{r^2}U,\qquad
H\longrightarrow-\frac1{r^2}V,\qquad
S_e\longrightarrow-\frac{2r_0}{r},\qquad
S_H\longrightarrow\frac1{r_0r}.
\label{eq:sstringfields}
\end{equation}
The sign of \(H\) follows from \(x=\cos\theta\), while its quadratic
contribution is sign-independent.
Equation~\eqref{eq:scopy} becomes
\begin{equation}
C_{\rm str}=\frac{r_0}{r^3}\bigl[\W(V)-\W(U)\bigr].
\label{eq:sstringcopy}
\end{equation}
We use the bivector operator convention
\((\widehat C\omega)_{ab}=\tfrac12C_{ab}{}^{cd}\omega_{cd}\), and order the independent pairs lexicographically as
\(01,02,03,04,12,13,14,23,24,34\).
The normalized operator is
\begin{equation}
\frac{r^3}{r_0}\widehat C_{\rm str}
=\operatorname{diag}
\left(1,-\frac12,-\frac12,0,-\frac12,-\frac12,0,1,0,0\right).
\label{eq:sstringspectrum}
\end{equation}
Each of \(\widehat{\W(U)}\) and \(\widehat{\W(V)}\) has rank ten,
while their difference has rank six. Every component involving index
4 vanishes. The four-dimensional kernel is the bivector subspace
\(T^*(\mathrm{Schwarzschild}_4)\wedge\dd Z\).
This is a cancellation between two nonzero projected squares.
The Schwarzschild string has De Smet type \(22\)
\cite{DeSmet2002}; the bivector rank alone is insufficient to establish
this spinor statement. Upon reduction along \(Z\), \(V=*_{4}U\) and
\(\W_4(V)=-\W_4(U)\), so the two five-dimensional sectors reduce to the
single Weyl-square direction of four-dimensional Schwarzschild.

\end{document}